\documentclass[a4paper,11pt]{article}
\pdfoutput=1 % to allow compilation in JCAP
 \usepackage{jcappub}
\usepackage{amsmath,amssymb}
\usepackage{bm}
\usepackage{xcolor}
\usepackage{graphicx}
\usepackage{enumitem}
\usepackage{geometry}
\usepackage{xspace}
\usepackage{pdflscape}
\usepackage{placeins}
\usepackage{epstopdf}
\usepackage{comment}
\usepackage{subfig}
\makeatletter
\gdef\@fpheader{}
\g@addto@macro\bfseries{\boldmath}
\makeatother

\newcommand{\ie}{{i.e.~}}

\newcommand{\OmegaPBH}{\Omega_{\mathrm{PBH}}}
\newcommand{\OmegaGW}{\Omega_{\mathrm{GW}}}
\newcommand{\rhoGW}{\rho_{\mathrm{GW}}}

\newcommand{\utot}{\mathrm{tot}}

\let\oldsqrt\sqrt
\def\sqrt{\mathpalette\DHLhksqrt}
\def\DHLhksqrt#1#2{%
\setbox0=\hbox{$#1\oldsqrt{#2\,}$}\dimen0=\ht0
\advance\dimen0-0.2\ht0
\setbox2=\hbox{\vrule height\ht0 depth -\dimen0}%
{\box0\lower0.4pt\box2}}

\newcommand{\dd}{\mathrm{d}}
\newcommand{\sss}[1]{{\scriptscriptstyle{#1}}}
\newcommand{\boldmathsymbol}[1]{{\ensuremath{\boldsymbol{#1}}}}

\newcommand{\uPl}{\mathrm{Pl}}

\newcommand{\umin}{\mathrm{min}}
\newcommand{\umax}{\mathrm{max}}

\newcommand{\urad}{\mathrm{rad}}

\newcommand{\usssPl}{\sss{\uPl}}

\newcommand{\ud}{\mathrm{d}}

\newcommand{\mPBH}{m_{\scriptscriptstyle{\mathrm{PBH}}}}
\newcommand{\rhoPBH}{\rho_{\scriptscriptstyle{\mathrm{PBH}}}}
\newcommand{\rhobarPBH}{\bar{\rho}_{\scriptscriptstyle{\mathrm{PBH}}}}

\newcommand{\deltaPBH}{\delta_{\scriptscriptstyle{\mathrm{PBH}}}}
\newcommand{\zetaPBH}{\zeta_{\scriptscriptstyle{\mathrm{PBH}}}}

\newcommand{\calH}{\mathcal{H}}
\newcommand{\calP}{\mathcal{P}}

\newcommand{\Mp}{M_\usssPl}

\newcommand{\beq}{\begin{equation}}
\newcommand{\eeq}{\end{equation}}
\newcommand{\bea}{\begin{equation}\begin{aligned}}
\newcommand{\eea}{\end{aligned}\end{equation}}

\newlength{\wsingfig}
\newlength{\wdblefig}
\newlength{\wquadfig}
\newlength{\wtriplefig}
\newcommand{\Eq}[1]{Eq.~(\ref{#1})}
\newcommand{\Eqs}[1]{Eqs.~(\ref{#1})}
\newcommand{\Fig}[1]{Fig.~{\ref{#1}}}

\newcommand{\Sec}[1]{Sec.~\ref{#1}}

\title{Gravitational waves induced from primordial black hole fluctuations: The effect of an extended mass function}
\author[a]{Theodoros Papanikolaou}
 
\affiliation[a]{National Observatory of Athens, Lofos Nymfon, 11852 Athens, 
Greece}

\emailAdd{papaniko@noa.gr}

\abstract{ 
The gravitational potential of initially Poisson distributed primordial black holes (PBH) can induce a stochastic gravitational-wave background (SGWB) at second order in cosmological perturbation theory. This SGWB was previously studied in the context of general relativity (GR) and modified gravity setups by assuming a monochromatic PBH mass function. Here we extend the previous analysis in the context of GR by studying the aforementioned SGWB within more physically realistic regimes where PBHs have different masses. In particular, starting from a power-law cosmologically motivated primordial curvature power spectrum with a running spectral index we extract the extended PBH mass function and the associated to it PBH gravitational potential which acts as the source of the scalar induced SGWB. At the end,  by taking into account the dynamical evolution of the PBH gravitational potential during the transition from the matter era driven by PBHs to the radiation era we extract the respective GW signal today.  Interestingly, in order to trigger an early PBH-dominated era and avoid the GW constraints at BBN we find that the running of the spectral index $\alpha_\mathrm{s}$ of our primordial curvature power spectrum should be within the narrow range $\alpha_\mathrm{s}\in[3.316,3.355]\times 10^{-3}$ while at the same time the GW signal is found to be potentially detectable by LISA.
}

\keywords{Scalar Induced Gravitational Waves, Primordial Black Holes}
\begin{document}

\maketitle
%%%%%%%%%%%%%%%%%% SECTION 1: INTRODUCTION  %%%%%%%%%%%%%%%%%%%%%%%%
\section{Introduction}
\label{sec:intro}
Primordial black holes (PBHs), firstly introduced in the early `70s~\cite{1967SvA....10..602Z, Carr:1974nx,1975ApJ...201....1C},  are formed in the early Universe before the period of star formation, out of the collapse of enhanced cosmological perturbations whose energy density is larger than a critical threshold~\cite{Harada:2013epa,Musco:2018rwt,Papanikolaou:2022cvo}. These 
objects have recently rekindled the interest of the scientific community since they can address a 
number of issues of modern cosmology.  They may indeed account for a part or the totality of dark matter
\cite{Chapline:1975ojl}, while additionally they can explain     
the large-scale structure formation through Poisson 
fluctuations \cite{Meszaros:1975ef,Afshordi:2003zb}.  Furthermore, they can 
potentially seed the supermassive black holes residing in the galactic centers~\cite{1984MNRAS.206..315C, Bean:2002kx},  while at the same time they can act as viable candidates for the progenitors of black-hole merging events~\cite{LIGOScientific:2018mvr}.  Other indications in favor of the PBH 
scenario can be found in~\cite{Clesse:2017bsw}.

On the other hand, given the huge progress in the field of GW astronomy there were numerous attempts connecting PBHs with gravitational waves~\cite{Sasaki:2018dmp}. 
Interestingly, one can connect PBHs with GWs associated to PBH merging events~\cite{Nakamura:1997sm, Ioka:1998nz, Eroshenko:2016hmn, Raidal:2017mfl, Zagorac:2019ekv,Hooper:2020evu}, with the SGWB of PBH Hawking radiated-gravitons~\cite{Anantua:2008am,Dong:2015yjs} as well as with GWs associated with an early matter-dominated era~\cite{Jedamzik:2010hq,Assadullahi:2009nf,Dalianis:2020gup} during which PBHs can be abundantly produced. In particular, within the last decade there has been witnessed an intense interest regarding the connection of PBHs with non linearly scalar induced gravitational waves (SIGW) generated at second order in cosmological perturbation theory~\cite{Bugaev:2009zh, Saito_2009, Nakama_2015, Yuan:2019udt,Fumagalli:2020nvq} given the fact that the same scalar perturbations that once enhanced collapse and form PBHs they can couple at second order with tensor perturbations and generate GWs [See~\cite{Domenech:2021ztg} for a recent review].

However, apart from the aforementioned connection channels of PBHs with GW 
signals, it has been recently noted in~\cite{Papanikolaou:2020qtd}, and 
further studied within GR~\cite{Domenech:2020ssp,Kozaczuk:2021wcl} and modified gravity setups~\cite{Papanikolaou:2021uhe,Papanikolaou:2022hkg}, that initially PBH Poisson isocurvature perturbations during a radiation-dominated (RD) era can be converted to adiabatic curvature perturbations in a subsequent PBH-dominated era and induce non-linearly a second order GW background at scales much larger than the PBH mean separation scale. 

Interestingly, it has been found that ultralight PBHs with masses $M<10^9\mathrm{g}$, evaporating before Big Bang Nucleosynthesis (BBN)~\cite{Kawasaki:1999na,Kawasaki:2000en,Hasegawa:2019jsa,Carr:2020gox}, can drive such early matter-dominated (eMD) eras and reheat the Universe through their evaporation~\cite{GarciaBellido:1996qt, Hidalgo:2011fj, Martin:2019nuw, Zagorac:2019ekv} while at the same time the associated to them SIGW background can be potentially detected by future GW experiments~\cite{Inomata:2020lmk,Domenech:2020ssp,Kozaczuk:2021wcl,Bhaumik:2022pil}. To the best of our knowledge, the SIGW background associated to PBH Poisson fluctuations  was studied within the context of monochromatic PBH mass functions with the transition from the PBH-dominated era to the RD era being well approximated as instantaneous~\cite{Inomata:2019ivs,Inomata:2020lmk,Domenech:2020ssp}.  However, one should go to more realistic regimes where PBHs are formed with different masses~\cite{1975ApJ...201....1C, Sureda:2020vgi,Ashoorioon:2019xqc,Ashoorioon:2020hln,Magana:2022cwq,Erfani:2021rmw} and see the effect on the respective SIGW background.

In this paper, we investigate this aspect by accounting for extended PBH mass functions.  In particular,  after giving in \Sec{sec:PBH_formation} the basics of the PBH formation formalism and computing the PBH mass function from a cosmologically motivated primordial curvature power spectrum we study in \Sec{sec:dynamical_Omega_PBH} the dynamical evolution of the background.  Followingly, in \Sec{sec:PBH_grav_potential} we extract the PBH gravitational potential, acting as the source of the SIGWs,  while in \Sec{sec:dynamical_PBH_grav_potential} we account for its dynamical evolution.  Then,  in \Sec{sec:SIGW} we deduce the respective GW signal by checking as well its potential detectability with LISA.  Finally, \Sec{sec:conclusions} is devoted to conclusions.

%%%%%%%%%%%%%%%%%  SECTION 2: PBH GAS %%%%%%%%%%%%%%%%%
\section{The primordial black hole formation formalism}\label{sec:PBH_formation}
\subsection{From the comoving curvature perturbation to the energy density contrast}\label{sec:delta_vs_zeta}
Assuming spherical symmetry on superhorizon scales, the overdensity region which collapses to form a black hole can be described by the following asymptotic form of the metric 
\beq
\mathrm{d}s^2 = -\mathrm{d}t^2 + a^2(t)e^{\zeta(r)}\left[\mathrm{d}r^2 + r^2\mathrm{d}\Omega^2\right],
\eeq
where $a(t)$ is the scale factor and $\zeta(r)$ is the comoving curvature perturbation which is conserved on superhorizon scales.  This curvature perturbation is directly related with the energy density contrast in the comoving gauge, the one standardly used in the PBH literature, though the following expression:
\begin{eqnarray}\label{eq:zeta_vs_delta:non_linear}
\frac{\delta\rho}{\rho_\mathrm{b}} &\equiv&\frac{\rho(r,t)-\rho_{\mathrm{b}}(t)}{\rho_{\mathrm{b}}(t)} = -\left(\frac{1}{aH}\right)^2\frac{4(1+w)}{5+3w}e^{-5\zeta(r)/2}\nabla^2e^{\zeta(r)/2},
\end{eqnarray}
where $H(t) = \dot{a}(t)/a(t)$ is the Hubble parameter and $w$ is the equation-of-state parameter defined as the ratio between the total pressure $p$ and the total energy density $\rho$, i.e. $w\equiv p/\rho$.  In the linear regime where $\zeta\ll 1$ \Eq{eq:zeta_vs_delta:non_linear} is reduced to
\beq\label{eq:zeta_vs_delta:linear}
\frac{\delta\rho}{\rho_\mathrm{b}}\simeq -\frac{1}{a^2H^2}\frac{2(1+w)}{5+3w}\nabla^2\zeta(r)  \Longrightarrow \delta_k =  -\frac{k^2}{a^2H^2}\frac{2(1+w)}{5+3w}\zeta_k.
\eeq
From the above relation, one sees that there is an one-to-one correspondence between $\zeta_k$ and $\delta_k$ meaning that statistical properties of $\zeta_k$ are inherited to $\delta_k$.  Thus, assuming a Gaussian statistical distribution for $\zeta_k$ as suggested from the observations entails the gaussianity of the energy density perturbations as well.  However,  the process of PBH formation is a non linear process and one expects that the amplitude of the critical threshold $\delta_\mathrm{c}$ to be in general non-linear. Consequently, one   one should consider the full non-linear relation between $\zeta$ and $\delta$ given by \Eq{eq:zeta_vs_delta:non_linear}.  At the end, one can show that the smoothed energy density contrast $\delta_\mathrm{m}$ is related with the linear Gaussian energy
density contrast $\delta_l$ through the following expression
~\cite{DeLuca:2019qsy,Young:2019yug}
\beq\label{eq:delta_m_smoothed}
\delta_\mathrm{m} = \delta_l - \frac{3}{8}\delta^2_l.
\eeq

At this point, we should highlight the fact that the use of $\zeta$ for the computation of the PBH mass function vastly overestimates the number of PBHs since scales larger than the PBH scale which are unobservable are not removed when the PBH distribution is smoothed~\cite{Young:2014ana}. Consequently, one should use $\delta$ instead given the $k^2$ damping on large scales as it can be seen by \Eq{eq:zeta_vs_delta:linear}.

\subsection{Smoothing the energy density perturbation}\label{sec:smoothing}

In order now to proceed to the extraction of the PBH mass function due to the gravitational collapse of non-Gaussian energy density perturbations, we make use of \Eq{eq:delta_m_smoothed} and we work with the linear Gaussian energy
density contrast denoted as $\delta_l$.  To do so, we smooth firstly the energy density contrast $\delta_l$ with a window function over scales smaller than the horizon scale avoiding in this way the formation of PBHs on small scales.  The large scales which remain unobservable are naturally removed due to the $k^2$ damping of $\delta$ as it can be seen in \Eq{eq:zeta_vs_delta:linear}.  Thus, one can write the smoothed energy density contrast as follows:
\beq
\delta^R_l = \int \mathrm{d}^3\vec{x}^\prime W(\vec{x},R)\delta(\vec{x}-\vec{x}^\prime) = \int_0^\infty4\pi r^2 W(r,R)\delta(r) \mathrm{d}r,
\eeq
where the first and the second equalities are in cartesian and spherical coordinates respectively and the function $W(\vec{x},R)$ is the smoothing window function. Throughout the paper, we will assume a volume-normalised Gaussian window function, whose Fourier transform is given by ~\cite{Young:2014ana}
\beq\label{eq:Gaussian_window_function}
\tilde{W}(R,k) = e^{-k^2R^2/2},
\eeq
with the smoothing scale $R$ being equal to the horizon scale, i.e. $R=(aH)^{-1}$. Regarding now the smoothed variance of the energy density field, it can be recast as 
\beq\label{eq:sigma}
\sigma^2 \equiv \langle \left(\delta^{R}_l\right)^2\rangle = \int_0^\infty\frac{\mathrm{d}k}{k}\mathcal{P}_{\delta_l}(k,R) = \frac{4(1+w)^2}{(5+3w)^2}\int_0^\infty\frac{\mathrm{d}k}{k}(kR)^4 \tilde{W}^2(k,R) \mathcal{P}_\zeta(k).
\eeq
where in the last equality we have used \Eq{eq:zeta_vs_delta:linear} and $\mathcal{P}_{\delta_l}(k,R)$ and $\mathcal{P}_{\zeta}(k)$ stand for the reduced energy density and curvature power spectra respectively. 

As regards the PBH mass, it is generally considered to be of the order of the horizon mass at PBH formation time, namely the horizon crossing time. More precisely, as it has been demonstrated in~\cite{Niemeyer:1997mt,Niemeyer:1999ak,Musco:2008hv,Musco:2012au},  the PBH mass spectrum should follow a critical collapse scaling law which reads as
\beq\label{eq:PBH_mass_scaling_law}
M_\mathrm{PBH} = M_\mathrm{H}\mathcal{K}(\delta-\delta_\mathrm{c})^\gamma,
\eeq
where $M_\mathrm{H}$ is the mass within the cosmological horizon at horizon 
crossing time, and $\gamma$ is the critical exponent which depends on the equation-of-state  parameter at the time of PBH formation. For radiation it is $\gamma\simeq 0.36$. The parameter $\mathcal{K}$ is a parameter that depends on the equation-of-state parameter and on the particular shape of the collapsing overdensity region. In the following we consider a representative value of $\mathcal{K}\simeq 4$, which corresponds to an initial 
Mexican-hat energy density perturbation profile with $\delta_\mathrm{c}\simeq 0.55$.

\subsection{Calculating the mass function within peak theory}\label{sec:beta_peak}
One then can proceed to the derivation of the PBH mass function $\beta(M)$.  In the following, we choose to work within the context of peak theory where the density of sufficiently rare and large peaks for a random Gaussian density field in spherical symmetry is given by~\cite{Bardeen:1985tr} 
\beq\label{eq:peak_density}
\mathcal{N}(\nu) = \frac{\mu^3}{4\pi^2}\frac{\nu^3}{\sigma^3}e^{-\nu^2/2},
\eeq
where $\nu \equiv \delta/\sigma$ and $\sigma$ is given by \Eq{eq:sigma}. The parameter $\mu$ is the first moment of the smoothed power spectrum given by 
\beq
\mu^2 =\int_0^\infty\frac{\mathrm{d}k}{k}\mathcal{P}_{\delta_l}(k,R)\left(\frac{k}{aH}\right)^2 = \frac{4(1+w)^2}{(5+3w)^2}\int_0^\infty\frac{\mathrm{d}k}{k}(kR)^4 \tilde{W}^2(k,R)\mathcal{P}_\zeta(k)\left(\frac{k}{aH}\right)^2.
\eeq

At the end, the fraction of the energy of the Universe at a peak of a given height $\nu$ which collapses to form a PBH, $\beta_\nu$ will be given by 
\beq
\beta_\nu = \frac{M_\mathrm{PBH}(\nu)}{M_\mathrm{H}}\mathcal{N}(\nu)\Theta(\nu - \nu_\mathrm{c})
\eeq
and the total energy fraction of the Universe contained in PBHs of mass $M$ reads as 
\beq\label{eq:beta_full_non_linear}
\beta(M) = \int_{\nu_\mathrm{c}}^{\frac{4}{3\sigma}}\mathrm{d}\nu\frac{\mathcal{K}}{4\pi^2}\left(\nu\sigma - \frac{3}{8}\nu^2\sigma^2 - \delta_{\mathrm{c}}\right)^\gamma \frac{\mu^3\nu^3}{\sigma^3}e^{-\nu^2/2},
\eeq
where $\nu_\mathrm{c} = \delta_{\mathrm{c},l}/\sigma$ and $\delta_{\mathrm{c},l}=\frac{4}{3}\left(1 -
\sqrt{\frac{2-3\delta_\mathrm{c}}{2}}\right)$.

Finally,  accounting for the expansion of the Universe which makes the PBH mass function increasing linearly with the scale factor during an RD era, the overall PBH abundance defined as $\Omega_\mathrm{PBH}\equiv \frac{\rho_\mathrm{PBH}}{\rho_\mathrm{tot}}$, where $\rho_\mathrm{tot}$ is the total energy density of the Universe, will be the integrated PBH mass function and will read as follows:
\beq\label{eq:Omega_PBH}
\Omega_\mathrm{PBH} (t) = \int_{M_\mathrm{min}}^{M_\mathrm{max}}  \left(\frac{M_\mathrm{H}(t)}{M}\right)^{1/2}\beta(M)\mathrm{d}\ln M,
\eeq
where $M_\mathrm{H}(t)$ is the mass within the cosmological horizon at time $t$. Note that in \Eq{eq:Omega_PBH} we have accounted for the fact that during the RD era $M_\mathrm{H}\sim a^{2}$ while no considering the effect of Hawking evaporation which is studied in the following sections.

\subsection{The primordial curvature power spectrum}
In the following, we choose to work with a  power-law  primordial curvature power spectrum $\mathcal{P}_\mathrm{\zeta}(k)$ which can be recast in the following form: 
\beq\label{eq:P_zeta_full}
\mathcal{P}_\mathrm{\zeta}(k) = A_\zeta\left(k/k_\mathrm{0}\right)^{n_\mathrm{s}(k)-1},
\eeq
with $A_\zeta$ being the amplitude of the comoving curvature perturbation power spectrum, $k_0$ a characteristic pivot scale and $n_\mathrm{s}(k)$ the scalar spectral index. 
The scale dependence of $n_\mathrm{s}(k)$ can be in general parametrised as follows~\cite{Kosowsky:1995aa}:
\beq
n_\mathrm{s}(k) = n_\mathrm{s,0} + \frac{\alpha_\mathrm{s}}{2!}\ln\left(\frac{k}{k_0}\right) - \frac{\beta_\mathrm{s}}{3!}\ln^2 \left(\frac{k}{k_0}\right) + ...
\eeq
In the following, we will consider the above expansion of $n_\mathrm{s}$ by considering up to $O\left[\ln\left(\frac{k}{k_0}\right)\right]$ terms. 

In the following,  given the fact that $A_\zeta$ and $n_\mathrm{s,0}$ are measured with a high statistical significance on the CMB scales, we normalise $A_\zeta$ and $n_\mathrm{s,0}$ in \eqref{eq:P_zeta_full} with respect to their values $A_\zeta=2.2\times 10 ^{-9}$  and $n_\mathrm{s,0}=0.9603$ on the CMB pivot comoving scale, namely $k_0=0.05\mathrm{Mpc}^{-1}$~\cite{Planck:2018vyg}.  Thus, in our physical setup the running of the spectral index $\alpha_\mathrm{s}$ is a free parameter of our cosmologically motivated power spectrum being strictly larger than $0$, i.e. $\alpha_\mathrm{s}>0$ in order to ensure an enhanced power on small scales where PBHs are assumed to be produced. 

One should stress here that one could assume as a first approximation that the spectral index $n_\mathrm{s}$ is scale-independent transforming it in this way to the free parameter of our physical setup.  However,  this would entail no running present for an incredibly large hierarchy of scales, namely from the CMB up to the PBH scales, which is rather questionable. For this reason,  in the following we consider as a first approximation the running of $n_\mathrm{s}$ the free parameter of the problem at hand.

\begin{figure}[t!]
\begin{center}
  \includegraphics[height = 8cm, width=10cm, clip=true]
                  {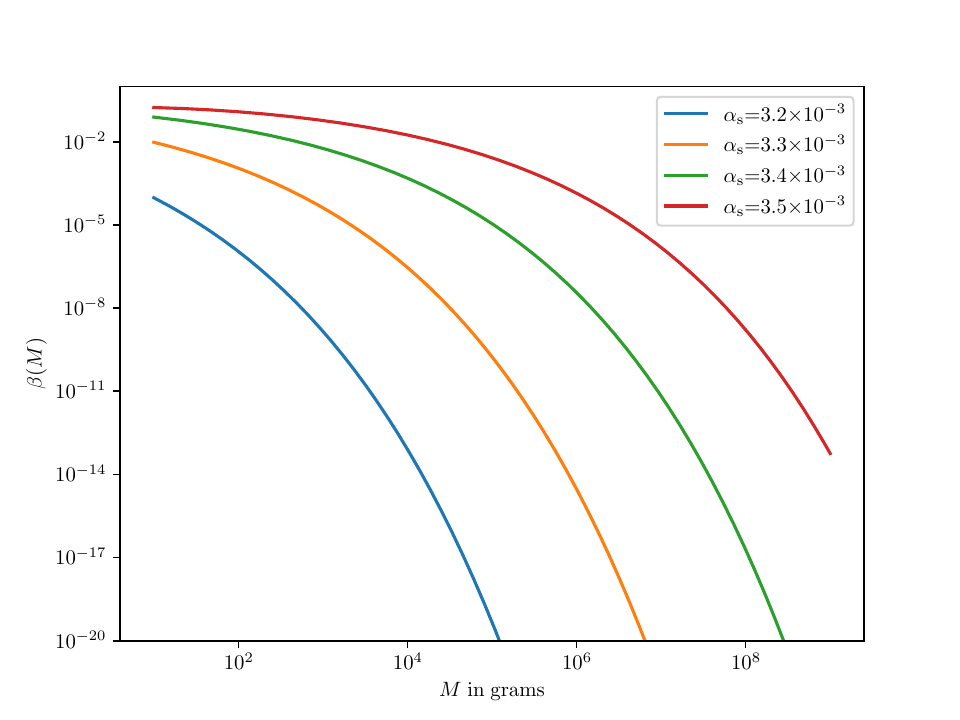}
  \caption{The PBH mass function for different values of the running of the spectral index $\alpha_\mathrm{s}$.
}
\label{fig:beta_alpha}
\end{center}
\end{figure}

Following therefore the procedure described in \Sec{sec:delta_vs_zeta}, \Sec{sec:smoothing}, \Sec{sec:beta_peak} one can derive the PBH mass function [See \Eq{eq:beta_full_non_linear}] which is shown in \Fig{fig:beta_alpha} within the range of masses we are interested in, namely $M\in[10\mathrm{g},10^9\mathrm{g}]$. 
As expected there is a higher probability for smaller mass PBHs to form since $\alpha_\mathrm{s}>0$. Interestingly,  we see as well that an increase in $\alpha_\mathrm{s}$ leads to an increase in the PBH mass function.  In particular, for $\alpha_\mathrm{s}>3.6\times 10^{-3}$ one is met with a PBH overproduction issue, that is that the PBH abundance becomes larger than $1$, namely $\OmegaPBH>1$. Consequently,  in the sections that follow we will restrict ourselves to $\alpha_\mathrm{s}\leq 3.6\times 10^{-3}$.  

%With this form for $\mathcal{P}_\mathrm{\zeta}(k)$ one can straightforwardly obtain the variance and the first moment of the smoothed energy density field  which read as 
%\beq
%\sigma^2 = \frac{8A_\mathrm{\zeta}}{81(k_0R)^{n_\mathrm{s}-1}}\Gamma\left(\frac{n_\mathrm{s}+3}{2}\right),\quad \mu^2 = \frac{n_\mathrm{s}+3}{2R^2}\sigma^2
%\eeq
%At the end, the PBH mass function reads as 
%\beq\label{eq:beta_ns}
%\beta(M) = \left(\frac{n_\mathrm{s}+3}{2}\right)^{3/2}\left\{\frac{\delta^2_\mathrm{c}}{2\sigma^2[R(M)]} + 2\right\}\frac{e^{-\frac{\delta^2_\mathrm{c}}{2\sigma^2[R(M)]}}}{4\pi^2},
%\eeq
%where the smoothing scale $R=(aH)^{-1}$ is directly related to the PBH mass by accounting for the fact that PBH mass is roughly equal to the horizon mass at horizon crossing time.  In particular, by demanding that $M= M_\mathrm{H}=\frac{4\pi}{3}\rho H^{-3}$ one can easily gets that
%\beq
%R(M) = \frac{M}{2a\pi\Mp^2}.
%\eeq
%

\section{The dynamical evolution of the background}\label{sec:dynamical_Omega_PBH}
In this work, we consider a gas of initially randomnly distributed in space PBHs with different masses at formation time within the range $M_\mathrm{f}\in[10\mathrm{g},10^9\mathrm{g}]$ which form during the RD era and evaporate before BBN~\cite{Kawasaki:1999na,Kawasaki:2000en,Hasegawa:2019jsa,Carr:2020gox}. Each black hole evaporates with its mass decreasing according to~\cite{Hawking:1974rv}
\begin{align}
M(t) = M_\mathrm{f}\left\lbrace
1-\frac{t-t_{\mathrm{f}}}{\Delta t_{\mathrm{evap}}(M_\mathrm{f})}\right\rbrace^{1/3}\, ,
\end{align}
with $t_{\mathrm{f}}$ being the PBH formation time and $\Delta t_{\mathrm{evap}}$ being the black hole evaporation time reading as~\cite{Hawking:1974rv}
 \beq\label{eq: t_evap}
 \Delta t_{\mathrm{evap}}(M_\mathrm{f}) =\frac{160}{\pi g_\mathrm{eff}}\frac{M^3_\mathrm{f}}{\Mp^4}, 
 \eeq
with $g_\mathrm{eff}$ being the effective number of degrees of freedom.  For the epochs we are interested in, namely before BBN  it can be approximated as 
$g_\mathrm{eff}\sim 100$ since it is the order of magnitude predicted by the Standard Model before the electroweak phase transition~\cite{Kolb:1990vq}.

Knowing therefore the dynamical evolution of the PBH masses, one can derive the dynamical evolution of the PBH abundance $\OmegaPBH (t)$ accounting for the effect of Hawking evaporation.  In particular,  if $\bar{\beta}$ denotes
the mass fraction in the absence of Hawking evaporation, one can write $\OmegaPBH (t)$ as
\begin{align}
\label{eq:OmegaPBH:continuous}
 \Omega_{\mathrm{PBH}}(t)  =
 \int_{M_\umin }^{M_\umax} \bar{\beta}
 \left(M,t\right)\left\lbrace
1-\frac{t-t_{\mathrm{ini}}}{\Delta t_{\mathrm{evap}}(M_\mathrm{f})}\right\rbrace^{1/3} \dd\ln M\, ,
\end{align}
where $t_{\mathrm{ini}}$ stands for the initial time  of our problem at hand which is actually the time of formation of the smallest mass PBH. The lower mass bound $M_\umin$ is given as the maximum between the minimum PBH mass at formation and the PBH mass which evaporates at time $t$ defined as $M_\mathrm{evap}(t)\equiv \left( \frac{\pi g_\mathrm{eff}\Mp^4 \Delta t_\mathrm{evap}}{160}\right) ^{1/3}$. Thus, one gets that  $M_\umin= \max [M_\mathrm{f,min},M_\mathrm{evap}(t)]$

Let us see now how  $\bar{\beta}$, which is the PBH mass function without accounting for the effect of Hawking evaporation,  can be derived. To do so,  we should consider the fact that the PBH energy density contained in an infinitesimal range of masses
$\delta (\ln M)$ is given by $ \delta \bar{ \rho} =
\rho_{\mathrm{tot}} \bar{\beta}\left(M,t\right) \delta (\ln M)
$. Since PBHs behave as matter with zero pressure, in the absence of Hawking
evaporation, one has that $\dot{\delta\bar{\rho}}+3H\delta\bar{\rho}
= 0$. Plugging the former expression into the latter, one gets that $
\left(\dot{\rho}_{\mathrm{tot}}+3 H {\rho}_{\mathrm{tot}}
\right)\bar{\beta}\left(M,t\right) + \rho_{\mathrm{tot}}
\dot{\bar{\beta}}\left(M,t\right) =0$.  Thus, one has that $\bar{\rho}_\utot =
\bar{\rho}_{\mathrm{PBH}} + \bar{\rho}_\urad$ and that in the absence of
Hawking evaporation,  $\dot{\bar{\rho}}_\utot = - 3 H
\bar{\rho}_{\mathrm{PBH}} - 4 H \bar{\rho}_\urad = -3H
\Omega_{\mathrm{PBH}}\bar{\rho}_\utot - 4 H (1-\Omega_{\mathrm{PBH}})
\bar{\rho}_\utot = H
\bar{\rho}_{\mathrm{tot}}(\Omega_{\mathrm{PBH}}-4)$. At the end, the equation for the dynamical evolution of $\bar{\beta}$ can be recast as:
\begin{align}
\label{eq:ode:beta:bar}
\dot{\bar{\beta}}(M,t)+H\left(\Omega_{\mathrm{PBH}}-1\right)\bar{\beta}(M,t)=0\, .
\end{align}
Since the above equation is linear and does
not depend on the PBH mass explicitly,  one can introduce the function $\mathfrak{b}$ that satisfies
\begin{align}
\label{eq:ode:frakb}
\dot{\mathfrak{b}}+H\left(\Omega_{\mathrm{PBH}}-1\right)\mathfrak{b}=0
\quad\text{with}\quad
\mathfrak{b}\left(t_{\mathrm{instab}}\right)=1\, ,
\end{align}
such as that $\bar{\beta}$ can be redefined as
\begin{align}
\label{eq:frakb:beta}
\bar{\beta}\left(M,t\right)
\equiv\bar{ \beta}\left(M,t_{\mathrm{f}}\right)\mathfrak{b}\left(t\right).
\end{align}
The set of equations~(\ref{eq:OmegaPBH:continuous}),
(\ref{eq:ode:frakb}) and~(\ref{eq:frakb:beta}) defines then a
system of coupled differential equations that one can integrate numerically.  At the end, the PBH and the radiation background energy densities will be given respectively by the following expressions:
\begin{eqnarray}\label{eq:rho_PBH+rho_r}
\rho_\mathrm{PBH}(t) & = & \OmegaPBH(t)\rho_\mathrm{tot}(t) \\
\rho_\mathrm{r}(t) & = & \left[1 - \OmegaPBH(t)\right]\rho_\mathrm{tot}(t),
\end{eqnarray}
where $\rho_\mathrm{tot} (t)$ is given by the Friedmann equation $\rho_\mathrm{tot} = 3\Mp^2H^2$.

\begin{figure}[t!]
\begin{center}
  \includegraphics[width=0.496\textwidth, clip=true]
                  {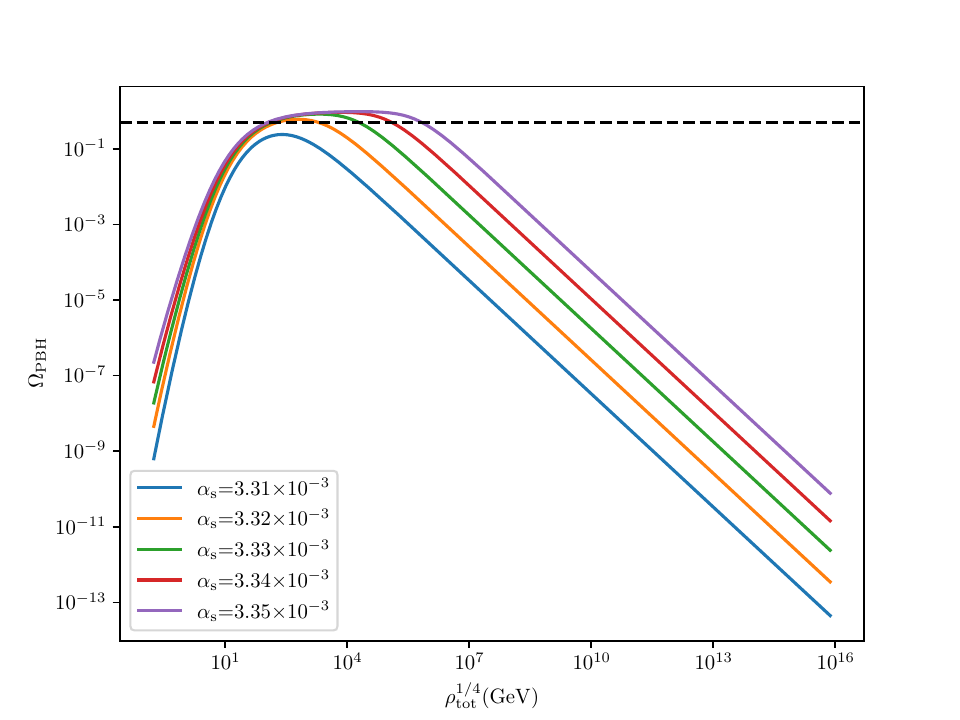}
  \includegraphics[width=0.496\textwidth, clip=true]
                  {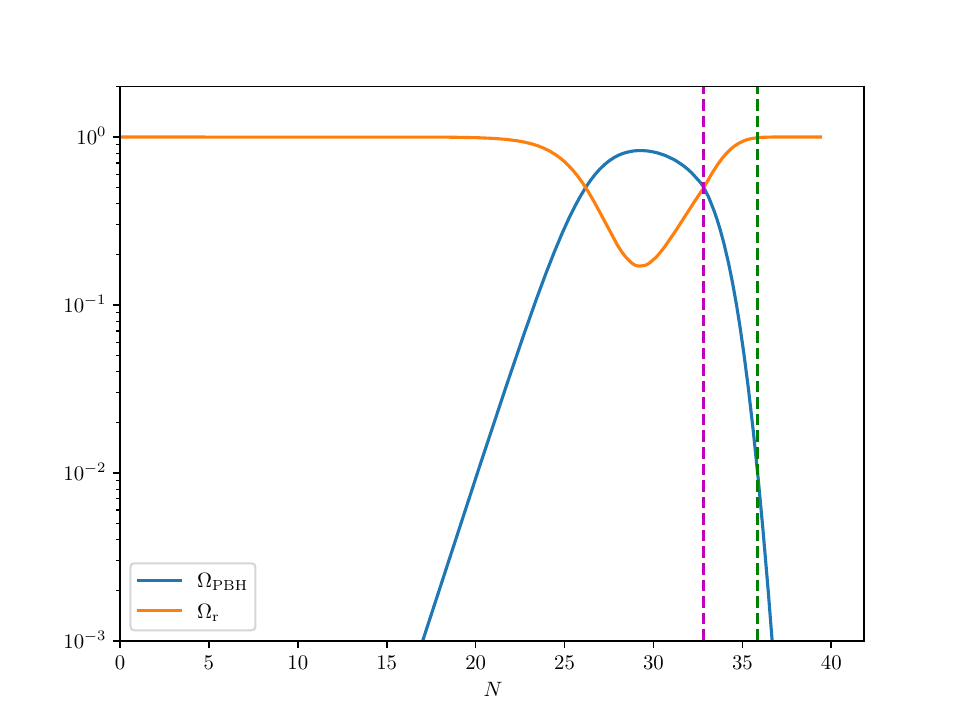}
  \caption{\textit{Left Panel: }The dynamical evolution of the background PBH energy density for different values of $\alpha_\mathrm{s}$. The dashed horizontal black line stands for transition point from the PBH to the radiation-dominated era, when $\OmegaPBH = \Omega_\mathrm{r} = 0.5$.  \textit{Right Panel:} The dynamical evolution of the background PBH and radiation energy densities in terms of the number of e-folds, $N$ for $\alpha_\mathrm{s}=3.33\times 10^{-3}$. The magenta vertical dashed line denotes the time of the onset of the radiation-dominated era, i.e. when $\Omega_\mathrm{r} = 0.5$ whereas the green dashed vertical line stands for the time when $\Omega_\mathrm{r} = 0.99$, namely when we are fully back to the radiation-dominated Universe. 
}
\label{fig:Omega_PBH_alpha}
\end{center}
\end{figure}

As one can see from the left panel of \Fig{fig:Omega_PBH_alpha}, the PBH abundance as we go from higher to lower energies, i.e. forward in time,  increases  initially linearly with the scale factor since initially when $\OmegaPBH\ll 1$ and Hawking evaporation is negligible,  $\Omega_\mathrm{PBH}=
\frac{\rho_\mathrm{PBH}}{\rho_\mathrm{tot}}\propto a^{-3}/a^{-4}\propto a$. Then at some point Hawking evaporation becomes important and $\Omega_\mathrm{PBH}$ starts its decreasing course.  Interestingly, all the curves start to decrease  roughly at the same time, which corresponds to the time at which the smallest PBH evaporates completely~\cite{Martin:2019nuw}. Furthermore, one may notice that as the running of the spectral index $\alpha_\mathrm{s}$ increases $\Omega_\mathrm{PBH}$  increases as well, a behavior which is expected since as we can see from \Fig{fig:beta_alpha}, an increase in $\alpha_\mathrm{s}$ is equivalent with an increase in the PBH mass function $\beta(M)$. 

In the right panel of \Fig{fig:Omega_PBH_alpha}, we see the evolution of $\Omega_\mathrm{PBH}$  and $\Omega_\mathrm{r}$ by choosing as our time variable the e-folds number $N$ defined as the logarithm of the scale factor $N=\ln a$.  As one may see, the e-folds passed between the onset of the RD era, i.e.  $\Omega_\mathrm{r} = 0.5$ (magenta dashed vertical line in the right panel of \Fig{fig:Omega_PBH_alpha}) and the time when we are fully back to the RD Universe, i.e. $\Omega_\mathrm{r} = 0.99$ (green dashed vertical line in the right panel of \Fig{fig:Omega_PBH_alpha}) is $\Delta N \sim 3 $, signaling that the transition between the PBH-dominated era to the RD era is gradual.  This phenomenology is kind of expected since now we are not in the monochromatic case. We have a gas of PBHs with different masses which evaporate at different times, hence rendering gradual the transition to the RD era.
\section{The primordial black hole gravitational potential}\label{sec:PBH_grav_potential}
In this section, following closely~\cite{Papanikolaou:2020qtd} we extract the power spectrum of the gravitational potential of a gas of PBHs with different masses.

\subsection{The matter power spectrum of PBH Poisson fluctuations}\label{sec:P_delta_Poisson}
We start by describing the statistics of the PBH density field.  To do so, we consider a gas of $N$ PBHs, each of them having a mass $m_{i}$ different from the mass of the other PBHs. We assume as well that PBHs are randomly distributed in space inside a volume $V$.  Thus, the location of each PBH is not correlated with the location of other PBHs, which is equivalent to consider that each black hole behaves as a point-like particle. In this way, we neglect the existence of an apparent horizon around the position of the centre of a black hole, which defines an exclusion zone.  However,  within this work we will not study scales smaller than the PBH apparent horizon which are not properly described in this setup, hence focusing on scales larger than the black hole size.

Let us now consider a sphere of radius $r$ (and of volume $v=4\pi r^3/3$) within the volume $V$, and denote with $P_{n}(r)$ the probability that $n$ PBHs are located inside this volume. For each PBH, the probability to be inside the sphere is given by $v/V$, and the probability to be outside is given by $(V-v)/V$, so one has
\bea
P_{n}(r) & = \binom{N}{n} \left(\frac{v}{V}\right)^n \left(1-\frac{v}{V}\right)^{N-n}\, .
\eea
By defining $\bar{r}$ as the mean distance between black holes, such  as that $V=4\pi \bar{r}^3 N/3$, the aforementioned probability can be recast as
\bea
\label{PDF for n PBHS inside r}
P_{n}(r) & =\binom{N}{n}\left(\frac{r^3}{N\bar{r}^3}\right)^{n}\left(1-\frac{ r^3}{N\bar{r}^3}\right)^{N-n}  
\underset{N\to \infty}{\longrightarrow}
\left(\frac{r}{\bar{r}}\right)^{3n}\frac{e^{-\frac{r^3}{\bar{r}^3}}}{n!},
\eea
where we have taken the large-volume limit. Such statistics are denoted as Poissonian. 

The total mass of PBHs contained inside the volume $v$ is then given by $n \langle \mPBH \rangle$ where $\langle M \rangle$ is the mean PBH mass defined through the PBH mass function as
\beq\label{eq:mean_PBH_mass}
\langle M \rangle (t) \equiv \frac{\int^{M_\mathrm{max}}_{M_\mathrm{min}}M \bar{\beta}
 \left(M,t\right)\left\lbrace
1-\frac{t-t_{\mathrm{ini}}}{\Delta t_{\mathrm{evap}}(M_\mathrm{f})}\right\rbrace^{1/3} \mathrm{d}\ln M}{\int^{M_\mathrm{max}}_{M_\mathrm{min}} \bar{\beta}
 \left(M,t\right)\mathrm{d}\ln M},
\eeq 
where in order to have the correct normalisation we have divided in the denominator with the PBH abundance without the effect of Hawking evaporation [See \Eq{eq:Omega_PBH} in the case of a RD dominated Universe].  
Thus,  the mean PBH energy density within the volume $v$ can be recast as 
\bea
\rhobarPBH(r) = \frac{n \langle M\rangle}{\frac{4}{3}\pi r^3}\, .  
\eea
Using now \Eq{PDF for n PBHS inside r}, one can obtain the two first moments of this quantity.  Therefore,  one firstly gets that
\bea
\label{eq:app:rho_bar}
\left\langle \rhobarPBH(r) \right\rangle = \sum_{n=0}^\infty P_n(r) \frac{n \langle M\rangle}{\frac{4}{3}\pi r^3} = \frac{\langle M\rangle}{\frac{4}{3}\pi \bar{r}^3},
\eea
corresponding to the average energy density.  Then,  proceeding to the second moment one finds that
\bea
\left\langle\rhobarPBH^2(r) \right\rangle = \sum_{n=0}^\infty P_n(r) \left(\frac{n \langle M\rangle}{\frac{4}{3}\pi r^3}\right)^2 =\frac{9\langle M\rangle^2}{16\pi^2 r^6}\left[\left(\frac{r}{
\bar{r}}\right)^{3} + \left(\frac{r}{\bar{r}}\right)^{6}\right]\, .
\eea
Combining the above expressions, one can find the variance of the energy density fluctuation,
\bea
\label{eq:rho2:1}
\left\langle \delta \rhobarPBH^2(r) \right\rangle =\left\langle \rhobarPBH^2(r) \right\rangle-\left\langle \rhobarPBH(r) \right\rangle^2=
\frac{9\langle M\rangle^2}{16\pi^2 \bar{r}^6}\left(\frac{\bar{r}}{r}\right)^{3}\, .
\eea

Followingly,  let us focus on the modeling of the PBH density field. As noted in~\cite{Papanikolaou:2020qtd} one can describe the gas of PBHs in terms of a fluid with an energy density $ \rhoPBH (\bm{x})$, and density contrast $\delta \rhoPBH (\bm{x})/\rho_\mathrm{tot}$, where $\delta \rhoPBH (\bm{x})= \rhoPBH (\bm{x}) - \left\langle \rhobarPBH\right\rangle$ and $\rho_\mathrm{tot}$ is the mean total energy density.  At the end,  $\rhobarPBH$ can be recast as
\bea
\rhobarPBH(r) &=\frac{1}{\frac{4}{3}\pi r^3} \int_{\vert \bm{x} \vert <r} \dd^3\bm{x}   \rhoPBH (\bm{x})\\
&=\left\langle \rhobarPBH \right\rangle + \frac{\rho_\mathrm{tot}}{\frac{4}{3}\pi r^3} \int_{\vert \bm{x} \vert <r} \dd^3\bm{x}   \frac{\delta  \rhoPBH (\bm{x})}{\rho_\mathrm{tot}} \, .
\label{eq:delta:rho:fluid}
\eea
Given the fact now that PBHs are randomly distributed in space, the position of PBH at location $\bm{x}$ is uncorrelated with the position of a PBH at location $\bm{x}'$, which entails that 
\bea
\label{eq:2pt:function:space:ansatz}
\left\langle \frac{\delta  \rhoPBH (\bm{x})}{\rho_\mathrm{tot}}\frac{\delta  \rhoPBH (\bm{x}')}{\rho_\mathrm{tot}}\right\rangle = \xi\,  \delta(\bm{x}-\bm{x}')\, ,
\eea
with $\xi$ depending a priori on $\bm{x}$ and $\bm{x}'$. However, given the statistical homogeneity and isotropy of the underlying density field,  $\xi$ depends only on $\vert \bm{x}-\bm{x}'\vert$. Thus, by taking the average of the square of \Eq{eq:delta:rho:fluid}, one can infer that
\bea
\label{eq:rho2:2}
\left\langle \delta \rhobarPBH^2(r) \right\rangle =\frac{9\rho_\mathrm{tot}^2}{16\pi^2r^6} \frac{4}{3}\pi r^3\, \xi\, .
\eea
By comparing  \Eqs{eq:rho2:1} and~\eqref{eq:rho2:2}, one can identify $\xi$ as
$\xi=3\langle M \rangle^2/(4 \pi\rho_\mathrm{tot}^2 \bar{r}^3)$. 
Then, by introducing the PBH fractional energy density $ \OmegaPBH =\left\langle \rhobarPBH \right\rangle/\rho_{\mathrm{tot}}$, \Eq{eq:2pt:function:space:ansatz} can thus be recast as 
\bea
\label{eq:2pt:function:real:space}
\left\langle \frac{\delta  \rhoPBH (\bm{x})}{\rho_\mathrm{tot}}\frac{\delta  \rhoPBH (\bm{x}')}{\rho_\mathrm{tot}}\right\rangle = \frac{4}{3}\pi \bar{r}^3  \OmegaPBH ^2\delta(\bm{x}-\bm{x}')\, ,
\eea
Going now to the comoving coordinates and defining the PBH density contrast with respect to the PBH energy density, which is the dominant component of the Universe during the eMD era driven by PBHs, one can rearrange the above formula as
\label{eq:2pt:function:comoving:space}
\bea
\left\langle \frac{\delta  \rhoPBH (\bm{x})}{\rho_\mathrm{PBH}}\frac{\delta  \rhoPBH (\bm{x}')}{\rho_\mathrm{PBH}}\right\rangle = \frac{4}{3}\pi \left(\frac{\bar{x}}{a}\right)^3  \delta(\bm{x}-\bm{x}')\, ,
\eea
where $\bm{x}$ stands for the comoving coordinates.  Finally, defining the power spectrum as $\langle \delta_{\bm{k}} \delta^{*}_{\bm{k}'} \rangle \equiv P_\delta(k) \delta(\bm{k}-\bm{k}')$ where $\delta_{\bm{k}}$ is the Fourier transform of $\deltaPBH$ one can obtain that
\bea\label{eq:delta_PBH_power_spectrum}
P_\delta(k)=\frac{4\pi}{3}\left(\frac{\bar{r}}{a}\right)^{3} \, .
\eea
As we can see from the above equation, $P_\delta(k)$ is independent of $k$,  as expected for Poissonian statistics.  At this point we should stress out that the description of the PBH gas in terms of a continuous fluid is only valid at scales larger than the PBH mean separation scale $\bar{r}$,  in order not to probe the granularity of the PBH density field and stay within the perturbative regime where $P_\delta(k)<1$.  This consideration imposes an ultra-violet cutoff defined as
\bea
\label{k_UV}
k_{\mathrm{UV}} \equiv \frac{a}{\bar{r}}, 
\eea
where $\bar{r}$ is given by
\beq\label{eq:rbar}
\bar{r}(t) = \left(\frac{3\langle M(t)\rangle}{4\pi\rho_\mathrm{PBH}(t)}\right)^{1/3}.
\eeq

At this point, one should highlight that the UV cutoff scale, usually denoted in the literature as exclusion scale, is  defined as the position of the maximum of the matter correlation function, which for large scales goes to zero.  Strictly speaking for the case of wide mass distributions this exclusion scale is not a single scale.  One in principle should run N-body simulations with compact objects of different masses and determine the exclusion scale as the position of the maximum of the matter correlation function, a calculation which has not been fully performed yet to the best of our knowledge.

Consequently, in our work not aiming to perform computationally high cost N-body simulations for the determination of the exclusion scale,  we focus on large scales imposing a UV cutoff scale related with the mean PBH separation scale defined in \Eq{eq:rbar}, an approximation which works quite well in the case of monochromatic mass functions (see here~\cite{Baldauf:2013hka} for the case of large scales structures).  

In our present case,  given the fact that the PBH mass function is tilted towards the lower masses (as it can be seen from \Fig{fig:beta_alpha}), the mean PBH mass at time $t$ will be tilted towards the lowest PBH mass present at that time.  Thus, the mean PBH separation scale \eqref{eq:rbar} will be tilted towards its lowest possible value\footnote{Note that in \Sec{sec:SIGW} the PBH mean separation scale and subsequently the PBH matter power spectrum is computed at PBH domination time and then is evolved through the use of the transfer function.  Thus, in \Eq{eq:rbar} $\rho_\mathrm{PBH}(t)$ is an increasing function of time up to PBH domination time.  } leading in this way to an underestimation of the PBH matter power spectrum.  In this sense, the GW amplitude and the bounds on the running of the spectral index $\alpha_\mathrm{s}$ derived in \Sec{sec:GW_spectrum} should be regarded as rather conservative.

One needs to comment as well on the possible seeding effect induced from the large mass PBHs.  In general, one expects that the cloud-in-cloud phenomenon, for which small mass primordial black holes may be absorbed by bigger mass ones, is actually absent due to the fact that the formation of a primordial black hole is an extremely rare event~\cite{MoradinezhadDizgah:2019wjf}. However, in our case as it can be seen from \Fig{fig:beta_alpha} we can have regimes where $\beta(M)$ can reach values up to $0.1$ so one can not treat PBH formation as a rare event.  In this dense PBH formation regime,  one should perform high cost numerical simulations within the excursion set formalism in order to extract the real PBH mass distribution~\cite{Auclair:2020csm}, something which is beyond the scope of this work. Interestingly, however, as it was shown in~\cite{Auclair:2020csm} for values of $\beta(M)<0.1$ the real PBH mass distribution differs slightly from the ``raw" PBH mass function [See \Eq{eq:beta_full_non_linear}].  The biggest change was observed in the region of small PBH masses wherever there was an extended plateau behavior of $\beta(M)$ at values of order $\mathcal{O}(1)$ - in particular $\beta(M)$ was reduced in this small mass region in order to account for the cloud-in-cloud phenomenon.  This feature however is absent in our case where the $\beta(M)$ profiles exhibit a far more abrupt decrease as we go to higher PBH masses \footnote{This behavior can be physically interpreted as follows: In the case where one is met with values of $\beta(M)$ of order $\mathcal{O}(1)$ for a large range of masses the cloud-in-cloud phenomenon is unavoidable.  In the case however, where $\beta(M)$ decreases abruptly as we go to higher PBH masses, which happens to be our case,  the cloud-in-cloud phenomenon is suppressed since heavier mass PBHs form with a dramatically lower probability compared to smaller mass PBHs. }. For values of $a_\mathrm{s}>3.6\times 10^{-3}$, one can observe such a plateau behavior, but for these high $a_\mathrm{s}$ values one is met with PBH overproduction already at formation time. Therefore,  as a first approximation, in what it follows the PBH seeding effect driven by heavier PBHs will be neglected. 

Concerning now accretion of the surrounding radiation, which would eventually increase the PBH mass,  recent analyses~\cite{DeLuca:2020bjf,DeLuca:2020fpg} find that within the regime of Bondi-Hoyle type accretion~\cite{1944MNRAS.104..273B} accretion is negligible when $m_\mathrm{PBH} < O(10)M_\odot$. Thus, in our case where we consider ultralight PBHs with masses smaller than $10^9\mathrm{g}$ the effect of accretion can be safely neglected.

\subsection{The power spectrum of the PBH gravitational potential}\label{sec:P_zeta}
We proceed now to the derivation of the power spectrum of the gravitational potential of the gas of PBHs,  denoted as $\Phi$, connecting it with the PBH matter power spectrum obtained in the previous section. To do so, we assume that PBHs are formed in the RD era which succeeds inflation. Considering now that PBHs are randomly distributed in space, their energy density is inhomogeneous while the total background energy density of radiation is homogeneous. Under these respects,  the PBH energy density perturbations can be viewed as isocurvature Poisson fluctuations with the associated Poissonian reduced power spectrum for the PBH density contrast being read as
\beq
\label{eq:reduced_P_delta}
\mathcal{P}_\delta(k) = \frac{k^3}{2\pi^2}P_\delta(k)= \frac{2}{3\pi} 
\left(\frac{k}{k_{\mathrm{UV}}}\right)^3 \Theta(k_\mathrm{UV}-k),
\eeq
where in the last equality we used \Eq{eq:delta_PBH_power_spectrum}.

Looking now into the dynamical evolution of the PBH abundance as shown in \Fig{fig:Omega_PBH_alpha}, $\OmegaPBH$ initially starts from small values and then increases reaching at some point the value $\OmegaPBH \simeq 1$. At this point,  the Universe enters a PBH-matter dominated era and as a consequence the initial isocurvature PBH perturbations during the RD era will be converted to adiabatic curvature perturbations in the subsequent PBH dominated era~\cite{Kodama:1986fg,Kodama:1986ud}.

Following now closely~\cite{Papanikolaou:2020qtd} we relate $\Phi$ and  $\delta_\mathrm{PBH}$ by introducing the 
uniform-energy density curvature perturbation for the PBH and radiation fluids~\cite{Wands:2000dp} defined as:
\bea
\zeta_\mathrm{r}=-\Phi+\frac14 \delta_\mathrm{r}
\eea
\bea
\zetaPBH=-\Phi+\frac13 \deltaPBH,
\eea
where $\Phi$ is the Bardeen potential~\cite{Bardeen:1980kt} as well as the isocurvature perturbation defined by 
\beq\label{eq:isocurvature_perturbation}
S = 3(\zetaPBH -\zeta_\mathrm{r}) = \delta_\mathrm{PBH} -\frac{3\delta_\mathrm{r}}{4}.
\eeq
At super-horizon scales $S$ is conserved,  thus it can be evaluated at PBH formation time. For the PBH scales which we are interested in, we ignore the radiation contribution and as a consequence $S$ can be identified with the PBH density contrast at formation time, i.e. $\deltaPBH(t_\mathrm{f})$. Focusing therefore on the PBH contribution one has that $\zeta \simeq \zetaPBH = \zeta_\mathrm{r} +S/3 \simeq S/3$.  Consequently,  one obtains for super-horizon scales that 
\bea
\label{eq:zeta:delta:superH}
\zeta\simeq \frac{1}{3} \delta_\mathrm{PBH}(t_\mathrm{f})\quad 
\mathrm{for}\quad 
k\ll \mathcal{H}\,. 
\eea
Thus,  during the PBH-matter dominated era, where $w=0$ and 
$\Phi$ is constant in time~\cite{Mukhanov:1990me}, using the fact that on 
super-horizon scales $\zeta \simeq - \mathcal{R}$~\cite{Wands:2000dp}, where 
$\mathcal{R}$ is the comoving curvature perturbation, as well as the relation 
between $\mathcal{R}$ and $\Phi$ in GR~\cite{Mukhanov:1990me}, we finally 
obtain 
that 
\cite{Papanikolaou:2020qtd,Papanikolaou:2021uhe}
\bea
\label{eq:Phi:delta:superH00}
\Phi\simeq -\frac{1}{5} \delta_\mathrm{PBH}(t_\mathrm{f})\quad 
\mathrm{for}\quad 
k\ll \mathcal{H}\,.
\eea
On the other hand, at sub-horizon scales the evolution of 
$\delta_
\mathrm{PBH}$ can be determined by solving the evolution growth equation for the matter density 
perturbations 
\begin{equation}
 \label{eq:growth:equation:GR}
\delta^{\prime\prime}_\mathrm{m}+\calH\delta^\prime_\mathrm{m}-4\pi 
G 
a^2\bar{\rho}_\mathrm{m}\delta_\mathrm{m} = 0 ,
\end{equation}
which, in the case of a Universe with radiation and PBH-matter, takes the form 
of the so-called M\'eszaros growth equation~\cite{Meszaros:1974tb}:
\bea\label{eq:Meszaros in GR}
\frac{\dd^2 \delta_\mathrm{PBH}}{\dd s^2}+\frac{2+3s}{2s(s+1)}\frac{\dd 
\delta_\mathrm{PBH}}{\dd s}-\frac{3}{2s (s+1)} \delta_\mathrm{PBH}=0\,.
\eea 
By solving the above equation we deduce that the dominant solution is given by 
\bea
\deltaPBH = \frac{2+3s}{2+3s_\mathrm{f}} \deltaPBH(t_\mathrm{f})\, .
\eea
It is important to highlight here that the aforementioned expression for the evolotion of $\deltaPBH$ is valid at all scales, and since it does not involve the  comoving scale $k$, it entails that the statistical distribution of PBHs remains of Poissonian type, \ie $\calP_{\deltaPBH}\propto k^3$ even after formation time~\cite{Desjacques:2018wuu, Ali-Haimoud:2018dau, MoradinezhadDizgah:2019wjf}.  Then,  deep in the PBH-dominated era, $\deltaPBH$ can be approximated as
~\cite{Papanikolaou:2020qtd}
\beq\label{eq:delta_PBH_sub_GR}
\delta_\mathrm{PBH}\simeq \frac{3s}{2}
\delta_\mathrm{PBH}(t_\mathrm{f}).
\eeq
Thus, given the fact that the Bardeen potential 
is related to the PBH density perturbation through the Poisson equation,  i.e.  $
\delta_\mathrm{PBH} =  -\frac{2}{3} \left(\frac{k}{\mathcal{H}}\right)^2\Phi$, one can insert \Eq{eq:delta_PBH_sub_GR} into the Poisson equation in order to find that 
\bea
\label{eq:Phi:delta:subH}
\Phi\simeq -\frac{9}{4}\left(\frac{k_{\mathrm{d}}}{k}\right)^2\, 
\delta_\mathrm{PBH}(t_\mathrm{f})
\quad \mathrm{for}\quad k\gg {\calH}_{\mathrm{d}}\, ,
\eea
with $k_\mathrm{d}\equiv \cal{H}_\mathrm{d}$ being the conformal Hubble parameter at the
PBH domination time.
Finally, making a crude interpolating between (\ref{eq:Phi:delta:subH}) and 
(\ref{eq:Phi:delta:superH00}),  and using (\ref{eq:reduced_P_delta}), we obtain that
\beq\label{eq:PowerSpectrum:Phi:PBHdom}
\mathcal{P}_\Phi(k) \equiv\frac{k^3}{2\pi^2}P_\Phi(k)= 
\frac{2}{3\pi}\left(\frac{k}{k_\mathrm{UV}}\right)^3 
\left(5+\frac{4}{9}\frac{k^2}{k_{\mathrm{d}}^2}\right)^{-2}\,.
\eeq

%
%%%%%%%%%%%%%%%%%%%%%%%%  SIGW %%%%%%%%%%%%%%%%%%%%%%%%%%%%%
\section{Evolving the primordial black hole  gravitational potential}\label{sec:dynamical_PBH_grav_potential}
In this section,  following~\cite{Inomata:2019zqy} we account for the evolution of the gravitational potential $\Phi$ during the transition from the PBH-dominated era to the RD era following PBH evaporation. This is quite crucial given the fact that $\Phi$ acts as the source of the SIGWs so the details of the dynamicall behavior of $\Phi$ during the transition will have a huge impact on the resultant GWs. 

Our physical system is comprised by matter in form of PBHs which ``decays" to radiation  through the process of PBH evaporation hence being similar to the decaying dark matter scenario. Consequently,  by making the correspondence $\rho_\mathrm{m}\longrightarrow \rho_\mathrm{PBH}$ and neglecting the anisotropic stress of radiation the perturbations can be described from the following system of equations in the Newtonian gauge\footnote{We choose to work in the Newtonian gauge as it is standardly used in the literature within the context of SIGWs.  The effect of the gauge dependence of the SIGWs is discussed in~\cite{Hwang_2017,Tomikawa:2019tvi,DeLuca:2019ufz,Yuan:2019fwv,Inomata:2019yww}}~\cite{Poulin:2016nat}:
\begin{align*}\label{eq:perturbations}
\delta^\prime_\mathrm{PBH} & = -\theta_\mathrm{PBH} + 3\Phi^\prime - a\Gamma\Phi\\
\theta^\prime_\mathrm{PBH} & = -\mathcal{H}\theta_\mathrm{PBH} + k^2\Phi \\ 
\delta^\prime_\mathrm{r} & = -\frac{4}{3}(\theta_\mathrm{r}-3\Phi^\prime) + a\Gamma\frac{\rho_\mathrm{PBH}}{\rho_\mathrm{r}}(\delta_\mathrm{PBH}-\delta_\mathrm{r}+\Phi) \\
\theta^\prime_\mathrm{r} & = \frac{k^2}{4}\delta_\mathrm{r} +k^2\Phi - a\Gamma\frac{3\rho_\mathrm{PBH}}{4\rho_\mathrm{r}}\left(\frac{4}{3}\theta_\mathrm{r} - \theta_\mathrm{PBH}\right),
\end{align*}
where $\delta_\alpha$ is the energy density contrast defined as $\delta_\alpha \equiv (\rho_\alpha - \rho_\mathrm{tot})/\rho_\mathrm{tot}$ with $\alpha=\left\{\mathrm{PBH},\mathrm{rad}\right\}$ and $\theta$ is the velocity divergence for each fluid component defined as $\theta\equiv \partial v_i/\partial x_i$ with $v_i$ being the fluid velocity while primes denote differentiation with respect to the conformal time $\eta$ defined as $\mathrm{d}\eta \equiv\mathrm{d}t/a$.

In the above expressions, $\Gamma$ is the PBH evaporation rate.  Given the fact that in our problem at hand we have different PBH masses evaporating at different times, one can define a mean PBH evaporation rate defined as
\beq\label{eq:mean_gamma}
\langle \Gamma\rangle(t) = \frac{\int_{t_\mathrm{evap,min}}^{t_\mathrm{evap,max}}\beta(t_\mathrm{evap})\Gamma_M(t_\mathrm{evap},t)\mathrm{d}\ln t_\mathrm{evap}}{\int_{t_\mathrm{evap,min}}^{t_\mathrm{evap,max}}\beta(t_\mathrm{evap})\mathrm{d}\ln t_\mathrm{evap}},
\eeq
where the PBH evaporation rate of a PBH with mass $M$, $\Gamma_M(t_\mathrm{evap},t)$ reads as~\cite{Hawking:1974rv,Inomata:2020lmk}
\beq\label{eq:gamma_M}
\Gamma_M(t_\mathrm{evap},t) \equiv -\frac{1}{M}\frac{\mathrm{d}M}{\mathrm{d}t} = \frac{1}{3(t_\mathrm{evap}-t)},
\eeq
with $t_\mathrm{evap}$ being a function of the PBH mass given by \Eq{eq: t_evap}.  The distribution function of the PBH evaporation times can be constructed by the PBH mass function at formation and accounting for the fact that $t_\mathrm{evap}\propto M^3$. At the end, considering conservation of probability, i.e. $\beta(M_\mathrm{f})\mathrm{d}\ln M_\mathrm{f} = \beta(t_\mathrm{evap})\mathrm{d}\ln t_\mathrm{evap}$, one can straightforwardly show that
\beq
\beta(t_\mathrm{evap}) = \beta\left[M_\mathrm{f}(t_\mathrm{evap})\right].
\eeq

The time evolution of $\Phi$ can be extracted from the Poisson equation and is given by~\cite{Mukhanov:2005sc}:
\beq\label{eq:Phi_evolution}
\Phi^\prime = -\frac{k^2\Phi + 3\mathcal{H}^2\Phi + \frac{3}{2}\mathcal{H}^2\left(\frac{\rho_\mathrm{PBH}}{\rho_\mathrm{tot}}\delta_\mathrm{PBH}+\frac{\rho_\mathrm{r}}{\rho_\mathrm{tot}}\delta_\mathrm{r}\right)}{3\mathcal{H}},
\eeq
while the time evolution of the background energy densities of PBHs and radiation is dictated by \Eq{eq:rho_PBH+rho_r}.

The above system of equations is solved numerically by adopting the following adiabatic initial conditions for the perturbations:
\beq\label{eq:initial_conditions}
\delta_\mathrm{PBH,ini}=-2\Phi_\mathrm{ini},\quad\delta_\mathrm{r,ini} = \frac{4}{3}\delta_\mathrm{PBH,ini},\quad\theta_\mathrm{PBH,ini} =\theta_\mathrm{r,ini} = 0.
\eeq
Restricting ourselves to the linear regime, the overall normalisation of the perturbations does not count. Thus, we take conventionally $\Phi_\mathrm{ini}=1$. 
\begin{figure}[t!]
\begin{center}
  \includegraphics[height = 8cm, width=10cm, clip=true]
                  {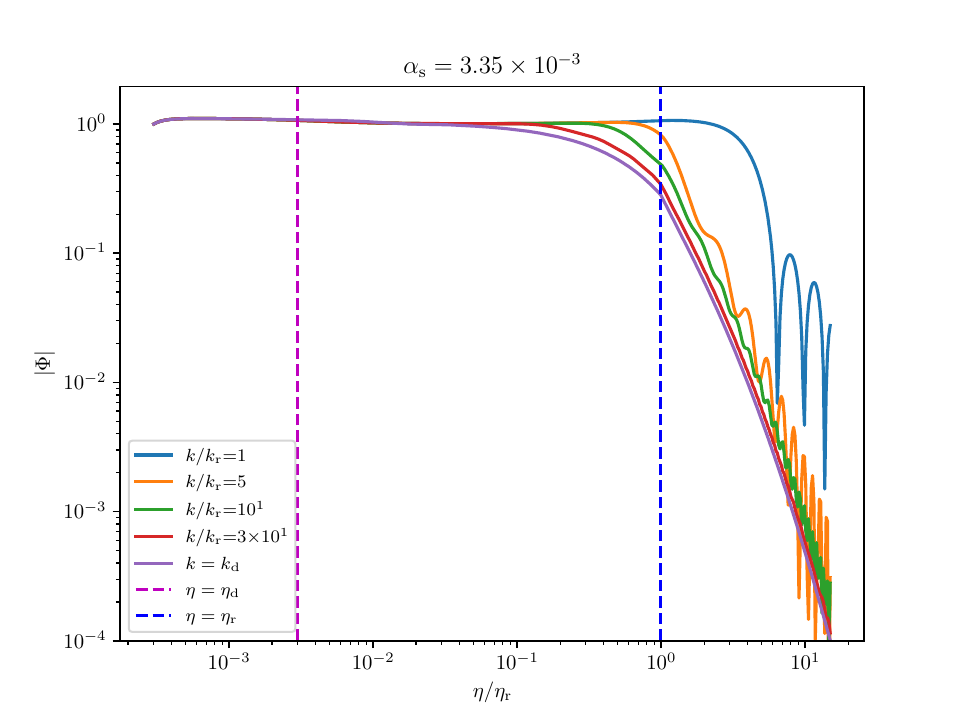}
  \caption{The dynamical evolution of $\Phi$ for different values of the comoving scales $k$ for $\alpha_\mathrm{s}=3.35\times 10^{-3}$.
}
\label{fig:Phi_alpha_0_00335}
\end{center}
\end{figure}

At this point, we should stress out that we work within the linear regime where matter perturbations should be less than unity.  During a matter-dominated era, like the one driven by PBHs,  the PBH energy density perturbation increases linearly with the scale factor, i.e. $\delta_\mathrm{PBH}\sim a$,  and for some scales $\delta_\mathrm{PBH}$ will become larger than one already before the end of the PBH-dominated phase. For this reason, one can define the non-linear scale, $k_\mathrm{NL}$ as the scale whose perturbation amplitude becomes unity at the onset of the RD era, i.e.  $\delta_{\mathrm{PBH}, k_\mathrm{NL}}(\eta_\mathrm{r}) = 1$. Given the fact now that $|\delta_{\mathrm{PBH},k}| \sim \mathcal{P}^{1/2}_\delta(k)$, where $ \mathcal{P}_\delta(k)$ is the reduced matter power spectrum defined in \Eq{eq:reduced_P_delta}, the PBH energy density pertrurbation at PBH domination time of a mode $k$ can be roughly estimated as 
\beq
\delta_k(\eta_\ud) \sim \sqrt{\frac{2}{3\pi}}\left(\frac{k}{k_\mathrm{UV}}\right)^{3/2}
\eeq
and starts growing linearly with the scale factor once the mode enters the horizon at a conformal time $\eta_k$. This is because the matter perturbations are frozen on super-Hubble scales and can only grow during the PBH-dominated era. One then  can solve $\delta_{\mathrm{PBH,} k_\mathrm{NL}}(\eta_\mathrm{r}) = 1$ and find that the non-linear scale can be recast after a straightforward calculation as
\beq\label{eq:k_NL}
k_\mathrm{NL} = k^{3/7}_\mathrm{UV}\left(\frac{3\pi}{2}\right)^{1/7}\left(\frac{a_\mathrm{d}}{a_\mathrm{r}}\right)^{2/7} \left(\frac{4a^2_\ud}{9t^2_\ud}\right)^{2/7}.
\eeq

Concerning now the smallest comoving scale considered in the computation of the SIGWs we want to be quite conservative choosing to work with only the scales which cross the horizon during the PBH-dominated era, namely $k_\mathrm{r}\leq k\leq k_\ud$~\cite{Inomata:2019zqy}. Therefore, the smallest scale considered, here or equivalently the larger comoving number,  will be given by 
\beq
k_\mathrm{max} = \min[k_\ud,k_\mathrm{NL}].
\eeq
At this point, it is very important to stress out that the PBH scales corresponding to PBH masses $M\in[10\mathrm{g},10^{9}\mathrm{g}]$ have entered the horizon deep in the early RD era before the onset of the PBH-dominated era.  At the end, the power spectrum of the gravitational potential $\Phi$ given in \Eq{eq:PowerSpectrum:Phi:PBHdom} will now read as:
\beq\label{eq:PowerSpectrum:Phi:PBHdom:k<k_d}
\mathcal{P}_\Phi(k) = 
\frac{2}{75\pi}\left(\frac{k}{k_\mathrm{UV}}\right)^3. 
\eeq

In \Fig{fig:Phi_alpha_0_00335}, we show for $\alpha_\mathrm{s}=3.35\times 10^{-3}$ the dynamical evolution of the gravitational potential $\Phi$ during the transition from the PBH-dominated era to the RD for different values of the comoving scale $k$. As one may, for modes crossing the horizon during the PBH era,  the gravitational potential initially decays and soon after the onset of the RD era it starts to oscillate due to the radiation pressure.  Interestingly, for values $k>5k_\mathrm{r}$ where $k_\mathrm{r}$ is the comoving scale crossing the Hubble radius at the onset of the RD era we see that gradually as we increase the value of $k$ the dynamical profile of $\Phi$ converges towards the dynamical profile of the mode crossing the Hubble radius at the onset of the PBH-dominated era, namely $\Phi_{k_\mathrm{d}}$.  The same behavior was confirmed as well for other values of the running of spectral index $\alpha_\mathrm{s}$. For this reason, when computing the SIGW signal in the next section we will consider only $\Phi_{k_\mathrm{d}}$ as the dynamical profile of $\Phi$ for all values of $k>5k_\mathrm{r}$ underestimating thus $\Phi$ at most by a factor of $2$ in the case where $k=5k_\mathrm{r}$ as it can be seen from \Fig{fig:Phi_alpha_0_00335}.

\section{Scalar induced gravitational waves}\label{sec:SIGW}
Having derived in the previous section the dynamical evolution of the gravitational potential $\Phi$ which actually seed the scalar induced gravitational waves in this section we move on computing the respective SGWB associated to the PBH Poisson fluctuations. 
\subsection{The tensor perturbations}\label{sec:tensor+perturbations}
Working in the Newtonian gauge and assuming zero anisotropic stress, the perturbed metric can be recast in the following form:
\bea
\label{metric decomposition with tensor perturbations}
\mathrm{d}s^2 = a^2(\eta)\left\lbrace-(1+2\Phi)\mathrm{d}\eta^2  + \left[(1-2\Phi)\delta_{ij} + \frac{h_{ij}}{2}\right]\mathrm{d}x^i\mathrm{d}x^j\right\rbrace,
\eea
where we have multiplied by a factor $1/2$ the second order tensor perturbation as is standard in the literature\footnote{The contribution from the first-order tensor perturbations is not considered in this work.}. Then, by performing a Fourier transform of the tensor perturbation and accounting for the two polarization states of the GWs in GR, the equation of motion for the tensor modes $h_\boldmathsymbol{k}$ reads as
\beq
\label{Tensor Eq. of Motion}
h_\boldmathsymbol{k}^{s,\prime\prime} + 2\mathcal{H}h_\boldmathsymbol{k}^{s,\prime} + k^{2} h^s_\boldmathsymbol{k} = 4 S^s_\boldmathsymbol{k}\, ,
\eeq
where  $s = (+), (\times)$. The source function $S^s_\boldmathsymbol{k}$ is given by
\beq
\label{eq:Source:def}
S^s_\boldmathsymbol{k}  = \int\frac{\mathrm{d}^3 \boldmathsymbol{q}}{(2\pi)^{3/2}}e^s_{ij}(\boldmathsymbol{k})q_iq_j\left[2\Phi_\boldmathsymbol{q}\Phi_\boldmathsymbol{k-q} + \frac{4}{3(1+w_\mathrm{tot})}(\mathcal{H}^{-1}\Phi_\boldmathsymbol{q} ^{\prime}+\Phi_\boldmathsymbol{q})(\mathcal{H}^{-1}\Phi_\boldmathsymbol{k-q} ^{\prime}+\Phi_\boldmathsymbol{k-q}) \right],
\eeq
The polarization tensors  $e^{s}_{ij}(k)$ are defined as~\cite{Espinosa:2018eve}
\beq
e^{(+)}_{ij}(\boldmathsymbol{k}) = \frac{1}{\sqrt{2}}
\begin{pmatrix}
1 & 0 & 0\\
0 & -1 & 0 \\ 
0 & 0 & 0 
\end{pmatrix}, \quad
e^{(\times)}_{ij}(\boldmathsymbol{k}) = \frac{1}{\sqrt{2}}
\begin{pmatrix}
0 & 1 & 0\\
1 & 0 & 0 \\ 
0 & 0 & 0 
\end{pmatrix}.
\eeq

One now can proceed by writing the Fourier transform of $\Phi$, i.e.  $\Phi_\boldmathsymbol{k}(\eta)$,  as $\Phi_\boldmathsymbol{k}(\eta) = \Phi(\eta) \phi_\boldmathsymbol{k}$, where $\phi_\boldmathsymbol{k}$ is the value of the gravitational potential at some initial time (which here we consider it to be the time at which PBHs dominate the energy content of the Universe,  $\eta_\ud$) and $\Phi(\eta)$ is the gravitational potential transfer function being actually the solution of \Eq{eq:Phi_evolution}. Consequently, \Eq{eq:Source:def} can be written in a compact form as
\beq
\label{Source}
S^s_\boldmathsymbol{k}  =
\int\frac{\mathrm{d}^3 q}{(2\pi)^{3/2}}e^{s}(\boldmathsymbol{k},\boldmathsymbol{q})F(\boldmathsymbol{q},\boldmathsymbol{k-q},\eta)\phi_\boldmathsymbol{q}\phi_\boldmathsymbol{k-q}\, ,
\eeq
where
\bea
\label{F}
\!\!\!\!\!
F(\boldmathsymbol{q},\boldmathsymbol{k-q},\eta) & \equiv 2\Phi(q\eta)\Phi\left(|\boldmathsymbol{k}-\boldmathsymbol{q}|\eta\right)  + \frac{4}{3(1+w)}\left[\mathcal{H}^{-1}\Phi^{\prime}(q\eta)+\Phi(q\eta)\right]
\\  & \kern-2em
\ \ \ \ \ \ \ \ \ \ \  \ \ \ \ \ \  \ \ \ \ \ \ \ \ \ \ \ \ \ \ \ \ \ \ \ \ 
\cdot \left[\mathcal{H}^{-1}\Phi^{\prime}\left(|\boldmathsymbol{k}-\boldmathsymbol{q}|\eta\right)+\Phi\left(|\boldmathsymbol{k}-\boldmathsymbol{q}|\eta\right)\right],
\eea
where the prime denotes differentiation with respect to the conformal time,  that is,  $\Phi^{\prime}(q\eta)\equiv \partial \Phi(q\eta)/\partial\eta = q \partial \Phi(q\eta)/\partial(q\eta)$.
The contraction  $e^s_{ij}(\boldmathsymbol{k})q_iq_j \equiv e^s(\boldmathsymbol{k},\boldmathsymbol{q})$ can be expressed in terms of the spherical coordinates $(q,\theta,\varphi)$ of the vector $\bm{q}$ as 
\beq
e^s(\boldmathsymbol{k},\boldmathsymbol{q})=
\begin{cases}
\frac{1}{\sqrt{2}}q^2\sin^2\theta\cos 2\varphi \mathrm{\;for\;} s= (+)\\
\frac{1}{\sqrt{2}}q^2\sin^2\theta\sin 2\varphi  \mathrm{\;for\;} s= (\times) 
\end{cases}
\, .
\eeq
At the end,  the tensor modes $h^s_\boldmathsymbol{k}$ can be obtained using the Green's function formalism where one can write $h^s_\boldmathsymbol{k}$ as
\bea
\label{tensor mode function}
h^s_\boldmathsymbol{k} (\eta)  =\frac{4}{a(\eta)} \int^{\eta}_{\eta_\mathrm{d}}\mathrm{d}\bar{\eta}\,  G^s_\boldmathsymbol{k}(\eta,\bar{\eta})a(\bar{\eta})S^s_\boldmathsymbol{k}(\bar{\eta}),
\eea
where the Green's function  $G^s_{\bm{k}}(\eta,\bar{\eta})$ is the solution of the homogeneous equation 
\beq
\label{Green function equation}
G_\boldmathsymbol{k}^{s,\prime\prime}(\eta,\bar{\eta})  + \left( k^{2} -\frac{a^{\prime\prime}}{a}\right)G^s_\boldmathsymbol{k}(\eta,\bar{\eta}) = \delta\left(\eta-\bar{\eta}\right),
\eeq
with the boundary conditions $\lim_{\eta\to \bar{\eta}}G^s_\boldmathsymbol{k}(\eta,\bar{\eta}) = 0$ and $ \lim_{\eta\to \bar{\eta}}G^{s,\prime}_\boldmathsymbol{k}(\eta,\bar{\eta})=1$.  
Below, we will use the Green's functions during the transition from a MD era to an RD era as well as during an RD era. Following closely~\cite{Kohri:2018awv,Inomata:2019zqy} one can show that 
\beq
kG_k^{\mathrm{MD}\rightarrow\mathrm{RD}}(x,\bar{x})=C(x,x_\mathrm{r})\bar{x}j_1(\bar{x}) + D(x,x_\mathrm{r})\bar{x}y_1(\bar{x}), \quad kG_k^{\mathrm{RD}}(x,\bar{x}) =\sin(x-\bar{x}),
\eeq
where $j_1$ and $y_1$ are the first and second spherical Bessel functions and the coefficients $C$ and $D$ read as
\beq
C(x,x_\mathrm{r}) = \frac{\sin x - 2x_\mathrm{r}(\cos x + x_\mathrm{r}\sin x) + \sin(x - 2x_\mathrm{r})}{2x^2_\mathrm{r}},
\eeq
\beq
D(x,x_\mathrm{r}) = \frac{(2x^2_\mathrm{r} -1)\cos x - 2x_\mathrm{r}\sin x + \cos(x-2x_\mathrm{r})}{2x^2_\mathrm{r}}
\eeq
by requiring continuity of the Green function $kG_k^{\mathrm{MD}\rightarrow\mathrm{RD}}(x,\bar{x})$ and its derivative at the transition time $\bar{x}=x_\mathrm{r}$.
\subsection{The gravitational-wave spectrum}\label{sec:GW_spectrum}
Having derived above the tensor perturbations,  we derive below the gravitational-wave spectrum.  To do so, we extract firstly  the tensor power spectrum, $\mathcal{P}_{h}(\eta,k)$  which is defined through the following relation:
\bea\label{tesnor power spectrum definition}
\langle h^r_{\boldmathsymbol{k}}(\eta)h^{s,*}_{\boldmathsymbol{k}^\prime}(\eta)\rangle \equiv \delta^{(3)}(\boldmathsymbol{k} - \boldmathsymbol{k}^\prime) \delta^{rs} \frac{2\pi^2}{k^3}\mathcal{P}^s_{h}(\eta,k),
\eea
where $s=(\times)$ or $(+)$.
Thus, after a straightforward but rather long calculation one acquires that $\mathcal{P}_{h}(\eta,k)$ for the $(\times)$ and $(+)$ polarization states can be recast as ~\cite{Ananda:2006af,Baumann:2007zm,Kohri:2018awv,Espinosa:2018eve} 
\bea
\label{Tensor Power Spectrum}
\mathcal{P}^{(s)}_h(\eta,k) = 4\int_{0}^{\infty} \mathrm{d}v\int_{|1-v|}^{1+v}\mathrm{d}u \left[ \frac{4v^2 - (1+v^2-u^2)^2}{4uv}\right]^{2}I^2(u,v,x)\mathcal{P}_\Phi(kv)\mathcal{P}_\Phi(ku)\,,
\eea
The two auxiliary variables $u$ and $v$ are defined as $u \equiv |\boldmathsymbol{k} - \boldmathsymbol{q}|/k$ and $v \equiv q/k$, and the kernel function $I(u,v,x)$ is given by
\bea
\label{I function}
I(u,v,x) = \int_{x_\mathrm{d}}^{x} \mathrm{d}\bar{x}\, \frac{a(\bar{x})}{a(x)}\, k\, G^s_{k}(x,\bar{x}) F_k(u,v,\bar{x}).
\eea
In the above expressions, $x=k\eta$ and $F_{k}(u,v,\eta)\equiv  F(k ,|\boldmathsymbol{k}-\boldmathsymbol{q}|,\eta)$ given the fact that the function $F(\boldmathsymbol{q},\boldmathsymbol{k-q},\eta)$ depends only on the modulus of its first two arguments.  

In what follows,  by accounting for the transition from the PBH-dominated era to the RD era and the continuity of the scale factor and the Hubble parameter, $a$ and $H$ are given by the following expressions in terms of the cosmic and the conformal times : 
\beq\label{eq:scale_factor}
a(t) =
\begin{cases}
a_\mathrm{d}\left(\frac{t}{t_\mathrm{d}}\right)^{2/3}\quad\mathrm{for\;}t<t_\mathrm{r}\\
a_\mathrm{r}\left(\frac{t}{t_\mathrm{r}}\right)^{1/2}\quad\mathrm{for\;}t>t_\mathrm{r}
\end{cases}
\xrightarrow{\mathrm{d}t\equiv\mathrm{d}\eta}
a(\eta) =
\begin{cases}
a_\mathrm{d}\left(\frac{\eta - \eta_\mathrm{d}}{A_1} + 1\right)^{2}\quad\mathrm{for\;}\eta<\eta_\mathrm{r}\\
a_\mathrm{r}\left(\frac{\eta-\eta_\mathrm{r}}{A_2} + 1\right)\quad\mathrm{for\;}\eta>\eta_\mathrm{r}
\end{cases}
\eeq
\beq\label{eq:Hubble_parameter}
H(t) =
\begin{cases}
\frac{2}{3t}\quad\mathrm{for\;}t<t_\mathrm{r}\\
\frac{1}{2t}\quad\mathrm{for\;}t>t_\mathrm{r}
\end{cases}
\xrightarrow{\mathrm{d}t\equiv\mathrm{d}\eta}
\mathcal{H}(\eta) \equiv \frac{a^\prime}{a} = 
\begin{cases}
\frac{2}{\eta - \eta_\mathrm{d} +A_1}\quad\mathrm{for\;}\eta<\eta_\mathrm{r}\\
\frac{2A_2}{\eta_\mathrm{r} - \eta_\mathrm{d} +A_1}\frac{1}{\eta - \eta_\mathrm{r} +A_2}\quad\mathrm{for\;}\eta>\eta_\mathrm{r},
\end{cases}
\eeq
where $^\prime$ denotes differentiation with respect to the conformal time,  $A_1 = 3t_\ud/a_\ud$ and $A_2 = 2t_\mathrm{r}/a_\mathrm{r}$. Thus, the kernel function in \Eq{I function} for times within the RD era, i.e.  $x>x_\mathrm{r}$, can be recast as follows:
\beq\label{eq:I_function_expanded_RD}
\begin{split}
I(u,v,x) & = \int_{x_\mathrm{d}}^{x_\mathrm{r}}\mathrm{d}\bar{x}\left(\frac{\bar{x}-x_\ud +A_1k}{x_\mathrm{r}-x_\ud +A_1k}\right)^2\frac{A_2k}{x-x_\mathrm{r}+A_2k}kG_k^{\mathrm{MD}\rightarrow\mathrm{RD}}(x,\bar{x})F_k(u,v,\bar{x}) \\ &  + \int_{x_\mathrm{r}}^x\mathrm{d}\bar{x}\frac{\bar{x}-x_\mathrm{r}+A_2k}{x-x_\mathrm{r}+A_2k}kG_k^{\mathrm{RD}}(x,\bar{x})F_k(u,v,\bar{x}).
\end{split}
\eeq

At the end,  using the flat spacetime approximation and treating the tensor perturbations as freely propagating GWs one can find~\cite{Maggiore:1999vm} that the GW spectral abundance defined as the GW energy density per logarithmic comoving scale can be recast as~\cite{Kohri:2018awv}
\beq\label{Omega_GW}
\Omega_\mathrm{GW}(\eta,k)\equiv \frac{1}{\bar{\rho}_\mathrm{tot}}\frac{\mathrm{d}\rho_\mathrm{GW}(\eta,k)}{\mathrm{d}\ln k} = \frac{1}{24}\left(\frac{k}{\calH(\eta)}\right)^{2}\overline{\mathcal{P}^{(s)}_h(\eta,k)},
\eeq
where $s=(\times)$ or $(+)$.  

In order now to compute the GW spectral abundance at our present epoch, one must evolve 
$\Omega_\mathrm{GW}(\eta,k)$ from a reference conformal time $\eta_\mathrm{*}$, where GWs start propagating as free waves, up to today.  Doing so, one has that 
\beq
\OmegaGW(\eta_0,k) = \frac{\rhoGW(\eta_0,k)}{\rho_\mathrm{c}(\eta_0)} = 
\frac{\rhoGW(\eta_\mathrm{*},k)}{\rho_\mathrm{c}(\eta_\mathrm{*})}\left(\frac{
a_\mathrm{*}}{a_\mathrm{0}}\right)^4 
\frac{\rho_\mathrm{c}(\eta_\mathrm{*})}{\rho_\mathrm{c}(\eta_0)} = 
\OmegaGW(\eta_\mathrm{*},k)\Omega^{(0)}_\mathrm{r}\frac{\rho_\mathrm{r,*}
a^4_\mathrm{*}}{\rho_\mathrm{r,0}a^4_0},
\eeq
where we have taken into account that $\Omega_\mathrm{GW}\sim a^{-4}$, and 
where the  
index 
$0$ refers to the present time. Then, considering that the radiation energy 
density reads as $\rho_r = 
\frac{\pi^2}{15}g_{*\mathrm{\rho}}T_\mathrm{r}^4$ and that the temperature of 
the radiation bath $T_\mathrm{r}$, scales as $T_\mathrm{r}\propto 
g^{-1/3}_{*\mathrm{S}}a^{-1}$, one finds that 
\beq\label{Omega_GW_RD_0}
\Omega_\mathrm{GW}(\eta_0,k) = 
\Omega^{(0)}_r\frac{g_{*\mathrm{\rho},\mathrm{*}}}{g_{*\mathrm{\rho},0}}
\left(\frac{g_{*\mathrm{S},\mathrm{0}}}{g_{*\mathrm{S},\mathrm{*}}}\right)^{4/3}
\OmegaGW(\eta_\mathrm{*},k),
\eeq
where $g_{*\mathrm{\rho}}$ and $g_{*\mathrm{S}}$ stand for the energy and entropy relativistic degrees of freedom.  In order to choose the reference conformal time $\eta_\mathrm{*}$ after which GWs behave as free waves one should account for the fact that as it can be seen from \Fig{fig:Phi_alpha_0_00335} during the RD era, the gravitational potential, which is the source of the SIGWs, decays on subhorizon scales. Consequently,  $\Omega_\mathrm{GW}$ becomes constant after the gravitational potential has sufficiently decayed during RD. As noted in~\cite{Inomata:2019zqy}, $\eta_{*}\sim \mathcal{O}(1)\eta_r$. Here, being quite conservative we take it $\eta_\mathrm{*}=10\eta_\mathrm{r}$. 

In order to ensure an eMD phase driven by PBHs one should require on the one hand that $\eta_\mathrm{r}\geq \eta_\mathrm{d}$. Since now an increase of $\alpha_\mathrm{s}$ entails an increase of the PBH abundance by requiring that $\eta_\mathrm{r}\geq \eta_\mathrm{d}$ one can find a lower bound on $\alpha_\mathrm{s}$.  As it was found numerically, in order to have an eMD phase driven by PBHs, $\alpha_\mathrm{s}\geq 3.316\times 10^{-3}$. 

On the other hand, one needs to avoid the GW constraints at BBN as well.  In particular, one has that the constraint on the energy density contribution of GWs at BBN time is $\Omega_{\mathrm{GW,BBN}}<0.05$.\footnote{We used the fact that $\Omega_{\mathrm{GW,BBN}}<\frac{7}{8}\left(\frac{4}{11}\right)^{4/3}\Delta N_\mathrm{eff}\sim 0.05$~\cite{Caprini:2018mtu} accounting for the latest Planck 2018 upper bound on $\Delta N_\mathrm{eff}$, namely $\Delta N_\mathrm{eff}<0.2$~\cite{Planck:2018vyg}. }
At the end,  one can translate this upper bound on $\Omega_{\mathrm{GW,BBN}}$ on an upper bound on the running of the spectral index $\alpha_\mathrm{s}$ since as explained before an increase in $\alpha_\mathrm{s}$ signals an increase in the duration of the PBH-dominated phase and consequently an increase on the amplitude of the SIGWs associated to PBH Poisson fluctuations. Interestingly, we found numerically that in order to avoid GW overproduction at BBN we obtain that $\alpha_\mathrm{s}<3.355\times 10^{-3}$.  At the end, the observationally interesting window for $\alpha_\mathrm{s}$ can be read as follows:
\beq\label{eq:alpha_constraints_GW_section}
3.316\times 10^{-3}<\alpha_\mathrm{s}<3.355\times 10^{-3}.
\eeq
As one can see,  it is quite narrow signaling the very high sensitivity of the SIGW signal on the spectral index. 

\begin{figure}[t!]
\begin{center}
  \includegraphics[height = 8cm, width=10cm, clip=true]
                  {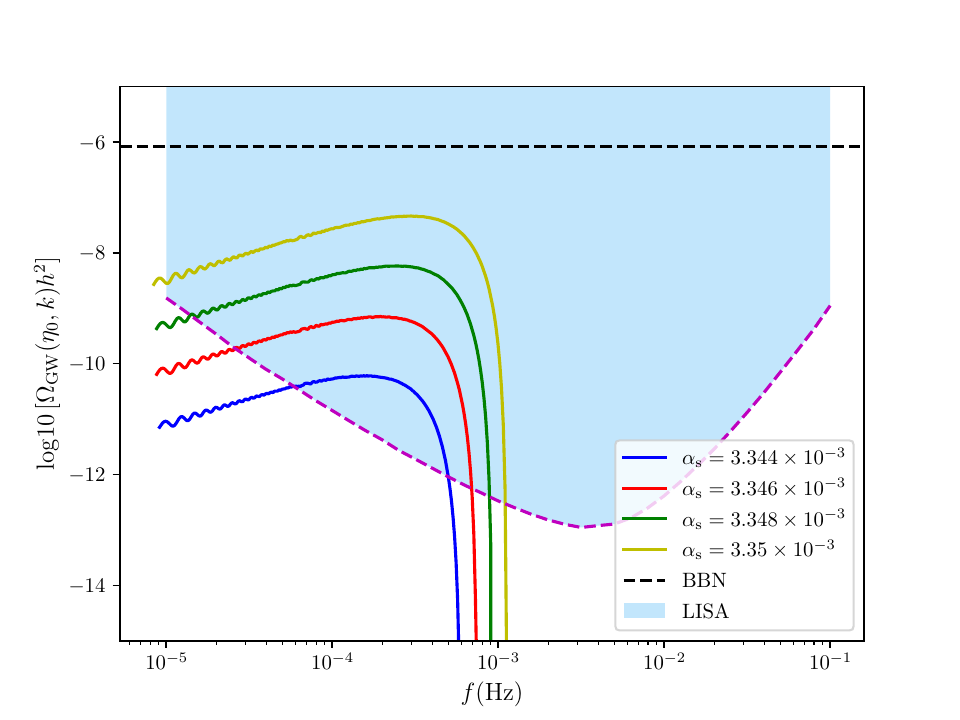}
  \caption{The scalar induced gravitational-wave spectrum for different values of the running of the spectral index $\alpha_\mathrm{s}$. With the magenta dashed curve we show also the LISA gravitational-wave sensitivity curve while with the black dashed horizontal line we show the upper limit on the amplitude of gravitational waves from BBN constraints.
}
\label{fig:GW_ns}
\end{center}
\end{figure}

This behavior can be verified in \Fig{fig:GW_ns} where we show in logarithmic scale the GW amplitude of the SIGWs associated to PBH Poisson fluctuations as a function of the frequency in $\mathrm{Hz}$ for different values of the running of the spectral index $\alpha_\mathrm{s}$ within the observationally interesting window \eqref{eq:alpha_constraints_GW_section}\footnote{As explained in \Sec{sec:dynamical_PBH_grav_potential} approximating $\Phi$ with $\Phi_{k_\mathrm{kd}}$ underestimates at most by a factor of $2$ its true value. Therefore, since from \Eq{Tensor Power Spectrum} and \Eq{Omega_GW}, $\OmegaGW\sim \Phi^4$,  in the GW signal curves shown in \Fig{fig:GW_ns} we have multiplied the GW spectral abundance with a factor $2^4=16$ accounting for the underestimation of the gravitational potential $\Phi$. }.  Astonishingly, an increase in the third digit after comma in $\alpha_\mathrm{s}$ is equivalent with an increase of a factor of $10$ in the amplitude of the GWs! We also find a characteristic oscillatory feature of the SIGW signal which is a characteristic signature of the gradualness of the transition from the PBH-dominated to the RD era as already noted in~\cite{Inomata:2019zqy}. 

Very interestingly,  we find as well that the frequency of the SIGWs studied here lies within the frequency band of LISA~\cite{Audley:2017drz} with the respective SIGW amplitude being orders of magnitude higher than the lowest GW amplitude detectable with LISA. Thus, the SIGW signal studied here,  constituting a characteristic signature of PBHs given also its oscillatory pattern, can be potentially detectable by LISA. 

%%%%%%%%%%%% CONCLUSIONS %%%%%%%%%%%%%%%%%%%%%%%%%%%%%
\section{Conclusions}\label{sec:conclusions}
In this paper, we have studied the scalar induced gravitational waves associated to the Poisson energy density fluctuations of ultralight PBHs with masses $M<10^9\mathrm{g}$ which evaporate before BBN and can potentially trigger an eMD era. During our analysis, we accounted for the effect of an extended PBH mass function, an aspect which was never studied before in the literature to the best of our knowledge. 

Following closely the analysis of~\cite{Papanikolaou:2020qtd} we studied carefully the PBH density field and extracted at the end the matter power spectrum of PBHs whose the initial spatial distribution follows Poisson statistics.  Then,  following~\cite{Inomata:2019zqy}, we studied the dynamical behavior of the background and the perturbations accounting carefully for the evolution of the PBH gravitational potential $\Phi$,  seeding the SIGWs,  during the gradual transition from the PBH-dominated era to the RD era. 

We focused mainly on the SIGWs which are sourced from the gravitational potential of modes $k$ which cross the Hubble radius during the eMD era driven by PBHs while at the same time we discarded the modes which become non-linear during the PBH-dominated era by imposing an appropriately chosen non-linear cutoff scale $k_\mathrm{NL}$.

At the end,  we applied our formalism for the computation of the SIGWs by adopting a cosmologically motivated power-law primordial curvature power spectum for the generation of the gas of PBHs. Interestingly, we found an oscillatory GW signal which is quite sensitive on the running of the spectral index $\alpha_\mathrm{s}$ of the power spectrum responsible for the PBH generation.  By requiring that PBHs can drive an eMD era and accounting as well for GW constraints from BBN we set an observational window for $\alpha_\mathrm{s}$, namely
\beq\label{eq:alpha_constraints}
3.316\times 10^{-3}<\alpha_\mathrm{s}<3.355\times 10^{-3}.
\eeq
Interestingly, we found that the frequency of the SIGWs studied here lies within the frequency band of LISA with the respective GW amplitude being orders of magnitude higher than the lowest GW amplitude detectable by LISA. Thus, the SIGW signal studied here,  constituting a characteristic signature of PBHs, can be potentially detectable by LISA. 

Here it is important to highlight that our formalism can be generalised to any type of primordial power spectrum responsible for the generation of PBHs triggering an eMD era either within GR~\cite{GarciaBellido:1996qt, Hidalgo:2011fj, Martin:2019nuw, Zagorac:2019ekv} or within modified gravity setups~\cite{Banerjee:2022xft,CANTATA:2021ktz}. In this way, one can use this portal as a novel probe to constrain on the one hand primordial power spectra on the small PBH scales which are not accesible by CMB or LSS probes~\cite{Kalaja:2019uju} and on the other hand gravitational theories responsible for PBH production.

Finally, we need to point out that our analysis can be extended in various ways.  In particular,  one can study as well the contribution from the modes entering the Hubble radius during the early RD era before the onset of the PBH-dominated phase as well GWs due to Hawking radiated gravitons~\cite{Domenech:2021wkk} which will clearly enhance the SIGW signal.

Additionally, given the fact that a large PBH abundance is reached when PBHs come to dominate the energy budget of the Universe, one expects the onset of PBH clustering which will obviously enhance the formation of PBH binaries~\cite{Raidal:2017mfl} as well as the formation of small scale virialised structures enhancing subsequently the power spectrum above its Poissonian value and leading in this way to an even larger GW signal than the one extracted in the present work. 

Given therefore all these considerations,  this work gives a lower bound on the potential GW signatures of the early PBH-domination scenario. The inclusion of all the extra aforementioned effects will clearly enhance the GW signal and subsequently tighten the constraints on $\alpha_\mathrm{s}$ hence rendering our results rather conservative.

%%%%%%%%%%%%%%%%%%%%%%%%  ACKNOWLEDGEMENTS %%%%%%%%%%%%%%%%%%%%%%%%%%%%%
\section*{Acknowledgments}
I would like to thank Vincent Vennin and Gabriele Franciolini for stimulating discussions and comments. I acknowledge financial support from the Foundation for Education and European Culture in Greece. 

%%%%%%%%%%%%%%%%%%%%%%%%  REFERENCES %%%%%%%%%%%%%%%%%%%%%%%%%%%%%
\bibliographystyle{JHEP} 
\bibliography{PBH_SIGW}
\end{document}